\documentclass[aps,twocolumn,showpacs,preprintnumbers,superscriptaddress]{revtex4-1}

\usepackage{array}
\newcolumntype{L}[1]{>{\raggedright\let\newline\\\arraybackslash\hspace{0pt}}m{#1}}
\newcolumntype{C}[1]{>{\centering\let\newline\\\arraybackslash\hspace{0pt}}m{#1}}
\newcolumntype{R}[1]{>{\raggedleft\let\newline\\\arraybackslash\hspace{0pt}}m{#1}}

\usepackage{amsmath}
\usepackage{amssymb}
\usepackage{amstext}
\usepackage{amsopn}
\usepackage{amsfonts}
\usepackage{amsxtra}
\usepackage[colorlinks]{hyperref}
\usepackage{url}
\usepackage{mathrsfs}
\usepackage[dvips]{graphicx}
\usepackage{relsize}
\usepackage{mathtools}
\usepackage{float}
\usepackage{bm}
\usepackage[dvipsnames]{color}
\numberwithin{equation}{section}

\newcommand{\mb}{\mathbf}

\newcommand{\h}{\hbar}

\newcommand{\ee}{\varepsilon}
\newcommand{\kk}{\mathbf{k}}

\newcommand{\ph}{\phantom\dag}

\newcommand{\vo}{\varrho}
\newcommand{\e}{\varepsilon}
\newcommand{\s}{\sigma}

\newcommand{\om}{\omega}
\newcommand{\al}{\alpha}

\newcommand{\de}{\delta}
\newcommand{\D}{\Delta}

\newcommand{\N}{\mathscr{N}}
\newcommand{\M}{\mathscr{M}}
\newcommand{\DD}{\mathscr{D}}
\newcommand{\Z}{\mathscr{Z}}

\newcommand{\G}{\mathscr{G}}
\def\mathclap#1{\text{\hbox to 0pt{\hss$\mathsurround=0pt#1$\hss}}}

\begin{document}
\def \brho{{\hbox{\boldmath $\rho$}}}
\def \beps{{\hbox{\boldmath $\epsilon$}}}
\def \bdelta{{\hbox{\boldmath $\delta$}}}

\title{Distinguishing the gapped and Weyl semimetal scenario in ZrTe$_5$: \\
insights from an effective two-band model}
\author{Z.~Rukelj }
\email[]{zrukelj@phy.hr}
\affiliation{Department of Physics, University of Fribourg, 1700 Fribourg, Switzerland}
\affiliation{Department of Physics, Faculty of Science, University of Zagreb, Bijeni\v{c}ka 32, HR-10000 Zagreb, Croatia}
\author{C.~C.~Homes}
\affiliation{Condensed Matter Physics and Materials Science Division, Brookhaven National Laboratory, Upton,
   New York 11973, USA}
\author{M.~Orlita}
\affiliation{LNCMI, CNRS-UGA-UPS-INSA, 25, avenue des Martyrs, F-38042 Grenoble, France}
\affiliation{Institute of Physics, Charles University in Prague, CZ-12116 Prague, Czech Republic}
\author{Ana Akrap}
\email[]{ana.akrap@unifr.ch}
\affiliation{Department of Physics, University of Fribourg, 1700 Fribourg, Switzerland}
\date{\today}

\begin{abstract}
Here we study the  static and dynamic  transport properties of a low energy two-band model proposed
previously in E. Martino et al. [PRL 122, 217402 (2019)], with an anisotropic in-plane linear momentum dependence,
and a parabolic out-of-plane dispersion.  The model is extended to include a negative band gap, which
leads to the emergence of a Weyl semimetal (WSM) state, as opposed to the gapped semimetal (GSM) state when
the band gap is positive. We calculate and compare the zero and finite frequency transport properties of
the GSM and WSM cases.
The $dc$ properties that are calculated for the GSM and WSM cases are Drude spectral weight, mobility and
resistivity.  We determine their dependence on the Fermi energy and crystal direction.  The in- and
out-of-plane optical conductivities are calculated in the limit of the vanishing interband relaxation rate
for both semimetals.  The main common features are  an $\om^{1/2}$ in-plane and $\om^{3/2}$ out-of-plane
frequency dependence of the optical conductivity.
We seek particular features related to the charge transport  that could unambiguously point to one ground
state over the other, based on the comparison with the experiment.  Differences between the WSM and GSM are
in principle possible only at extremely low carrier concentrations, and at low temperatures.
\end{abstract}

\maketitle

\section{Introduction}
Zirconium pentatelluride, ZrTe$_5$, is a layered material \cite{Okada1980, Jones1982, Whangbo1982, Shahi2018}
which recently became a topic of intense research.  This was mainly due to the experimental evidence
\cite{Wang2018, Zhang2017a, Yuan2016} of a 3D Dirac-like  band structure in the vicinity of the $\Gamma$ point
of the Brillouin zone, a novelty compared with the previously held belief of parabolic like valence bands \cite{Whangbo1982}.
One of the major signatures of a 3D Dirac-like band structure is the  linearity in the optical conductivity with respect to
photon energy $\h \om$  above the Pauli threshold \cite{AshbyPRB14}.  However, for $\rm{ZrTe_5}$ recent optical and magneto-optical
measurements \cite{Martino2019} suggest that the energy bands are not entirely linear, but posses an out-of-plane parabolic term
as well.

Like in many other topological semimetals, in ZrTe$_5$ the intrinsic energy scales are small. This makes it challenging
to experimentally distinguish between different possible ground states \cite{Moreschini2016}.  The ambiguity of the
bandgap --- whether it is zero, finite and positive, or finite and negative --- also opens a possibility that ZrTe$_5$
may be a Weyl semimetal \cite{ Liang2018}, and not Dirac semimetal as previously stated \cite{Chen2015, Chen2015m}.
To distinguish between these two options, it is of interest to see how much their calculated charge transport quantities
differ. This begs the question of whether one could interpret the same experimental data in different ways.

Based on the experiment and the {\it ab initio} calculation, we had previously introduced a simple low energy two-band
model for ZrTe$_5$, which was identified as  gapped semimetal.  The main features of the proposed effective
Hamiltonian are the gapped and electron-hole symmetric eigenvalues. This is accompanied by the anisotropic linearity
of the bands along the intralayer $(x,y)$ directions and the parabolic dispersion in the weakly dispersive out-of-plane
$z$ direction.  This model provided an explanation of experimental data \cite{Martino2019, Santos-Cottin2020}, in
particular the  square-root dependence of the optical conductivity at very low photon energies, in contrast to the
linear dependence found in  3D Dirac semimetals \cite{AshbyPRB14, Tabert2016, Tabert2016a}. It also allowed us to
estimate the energy interval in which the simple two-band model applies.

In this work, we identify under which circumstances it is possible to distinguish between the gapped and Weyl semimetal
scenario, specifically for ZrTe$_5$. To do this, we generalize the Hamiltonian model to allow for a negative bandgap
\cite{Mukherjee2019, Lu2015, Okugawa2014}. By this simple change of the sign of the bandgap, we generate a minimal
$2\times 2$ model Hamiltonian for Weyl semimetal. And so, by changing the sign of this parameter,  we pass from
a gapped semimetal (GSM) to the Weyl semimetal (WSM).  The main difference lies in the shape of the bands at
low energies. Contrary to the GSM case, the WSM case has a 3D linear-like bands in the close vicinity of the
two Weyl points.

In the $\om=0$  case, corresponding to $dc$ transport, we calculate the total and the effective concentration
of electrons. Since the effective concentration is direction dependent, it will explain the resistivity
anisotropy as well as the carrier mobility.

All the three spatial components of the real part of the interband conductivity are evaluated for GSM and WSM
cases, in the limit of vanishing relaxation rate.  We find that, for both the GSM and WSM, at high photon
energies the $(x,y)$ plane conductivity has a $(\h\om)^{1/2}$ dependence, and in the $z$ direction it has
a $(\h \omega)^{3/2}$ dependence when the external field energies are well above the bandgap value.
For photon energies below the bandgap, the GSM optical conductivity is zero, while the WSM shows a $\h\om$ dependence,
similar to the 3D Dirac case.

Finite temperatures, with $k_B T$ comparable to the Fermi energy, significantly alter the shape of the optical conductivity.
This results in a linear-like optical conductivity, which can easily be mistaken for a signature of a gapless
3D Dirac dispersion.
Finite interband relaxation values only slightly modify the general appearance of the real part of the
conductivity, except in the bandgap region where the conductivity acquires a finite contribution
proportional to the relaxation itself.

%
%
\section{{\em Ab initio} calculations and the model Hamiltonian}\label{hami}
\begin{figure}[b]
\centering
\hspace{-4mm} \includegraphics[width=6.9cm]{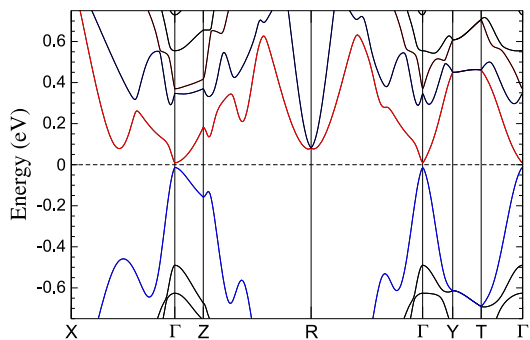}
\caption{{\it Ab initio} calculated band structure of $\rm{ZrTe_5}$. The valence and the conduction bands
are drawn in blue and red, respectively.}
\label{f1}
\end{figure}

We have performed the {\it ab initio}  band structure calculations of the orthorhombic $Cmcm$ phase of ZrTe$_5$
using density functional theory (DFT) with the generalized gradient approximation \cite{singh,singh91,wien2k}.
Once the unit cell is finalized with the parameters $a = 4.06$~\AA, $b = 14.76$~\AA, and $c = 13.97$~\AA, a
spin-orbit coupling is added to the electronic structure calculations. The results are shown in Fig.~\ref{f1}
with the valence band in blue, and the conduction band in red.

At small energies the material appears to be a semimetal with a small bandgap and the quasi-linear features
in the vicinity of the  $\Gamma$ point in the Brillouin zone.  The effective model considered in \cite{Martino2019}
is based on these basic features of the calculated band structure.
%
%
However, the problem lies in the values of the {\it ab initio} calculated parameters in Table \ref{tabla},
which  deviate from the experimentally determined parameters \cite{Martino2019}.  In particular, the
bandgap $2\D$ here is off by a factor of three, and in some references a factor of ten or
more \cite{Xiong2017, Miller2018, Fan2017}.

\begin{table}[t]
\caption{The values of the parameters entering Hamiltonian (\ref{ham1}). The parameters $\Delta$ and $\ee_F$ are
taken from the magneto-optical measurements \cite{Martino2019} while velocities and effective mass have been calculated based on the
comparison of the theoretical predictions of the GSM model and the experimental values.}
\vspace*{0.1cm}
\begin{ruledtabular}
\begin{tabular}{c||ccccc}
 & $v_x ({\rm{m/s}})$ & $v_y ({\rm{m/s}})$ &$ m^*/m_e$ & $2\D ({\rm{meV}})$ & $\e_F ({\rm{meV}})$ \\
 \cline{1-6}
 exp &${7\times10^5}^{\ph}$  & ${5\times10^5}^{\ph}$ & $2$ & $6$ & $14$ \\
 DFT &${3\times10^5}^{\ph}$  & ${2\times10^5}^{\ph}$ & $1$ & $20$ & $0$ \\
\end{tabular}
\end{ruledtabular}
\label{tabla}
\end{table}
%

%
%
\subsection{Effective two-band model}
The $2\times 2$ Hamiltonian matrix implements the electron-hole symmetry of the valence bands, a positive
energy band gap $2\D$ originating from the spin-orbit coupling, with the assumption of a free-electron like
behavior in the $z$ (or $b$ axis) direction and linear energy dependence in the $x,y$ ($a,c$) plane.
Here we expand the model to account for the Weyl phase by adding a negative bandgap. The Hamiltonian is thus
\begin{equation} \label{ham1}
  \hat{H}_\nu = \h v_x k_x\s_x + \h v_y k_y\s_y + (\h^2 c^2k_z^2 +\nu \D)\s_z,
\end{equation}
where the label $\nu$ differentiates between the GSM for the value $\nu = +1$, and the WSM for the value
$\nu = -1$.  Further, $\s_{x,y,z}$ are Pauli matrices, $v_{x,y}$ are the velocities in the $x$ and $y$
directions, and we introduce $c^2 = 1/2m^*$ with $m^*$ being the effective mass.

The diagonalization of Eq.~(\ref{ham1}) gives electron-hole symmetric eigenvalues
\begin{equation}\label{ham2}
  \ee^{c,v}_{\nu}(\kk) = \pm \sqrt{(\h v_xk_x)^2 + (\h v_yk_y)^2 + ( \h^2 c^2k_z^2 +\nu \D )^2 },
\end{equation}
with the indices for the conduction ($c$) and valence ($v$) bands.
Although trivial, the change from $\D \to -\D$ significantly alters the energies and single-particle properties.
While the GSM phase is always gapped in this model, the WSM phase has two Weyl points in the Brillouin zone
where the energy vanishes, $(k^w_x,k^w_y,k^w_z) = (0,0,\pm \sqrt{\D}/\h c)$.
Expanding the WSM eigenvalues around these two points gives  linear momentum eigenvalues,
\begin{equation}\label{ham0}
  \ee^{c,v}_W(\kk-\kk^w) = \pm \sqrt{(\h v_xk_x)^2 + (\h v_yk_y)^2 + (\h v_zk_z)^2},
\end{equation}
where we can formally identify
\begin{equation}\label{ham00}
  v_z^2 = 4 \D c^2 = 2\D/m^*.
\end{equation}%
In a third, trivial phase, a zero gap phase occurs when the bandgap is set to zero, $\D = 0$.

For $\D>0$, we have a gapped phase in which the gap is $k_z$-dependent, but it never changes its sign.
Therefore, there is no interesting topology involved \cite{Tchoumakov2017}.
In contrast, for $\D<0$, we obtain a minimal model for a  WSM. This model is spin degenerate simply
because the Hamiltonian matrix is $2\times2$ and not $4\times4$.
Spin degeneracy is ensured by the centrosymmetric lattice of ZrTe$_5$.
Still, because we gain in simplicity, it is fitting to call this $\D<0$ phase in a $2\times2$ Hamiltonian
model a Weyl semimetal phase \cite{Mukherjee2019, McCormick2017, SSQTI}.

%
%
\begin{figure}[t]
\centering
\hspace{-4mm} \includegraphics[width=6.9cm]{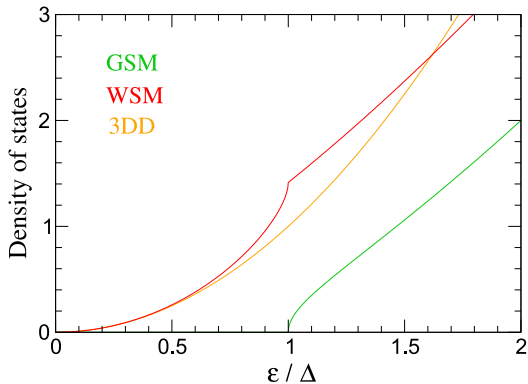}
\caption{The density of states (DOS) as a function of energy in units of $g_0$ [defined
in Eq.~(\ref{g555})] is shown for three cases: Weyl semimetal [Eq.~(\ref{g6})]; gapped semimetal
[Eq.~(\ref{g8})]; and 3D Dirac dispersion [Eq.~(\ref{g9})].  At $\e \ll \D$, the DOS for the Weyl
case and 3D Dirac dispersion coincide.  The DOS is plotted in units of $g_0\D^{3/2}$ as a
function of $\e/\D$.}\label{f2}
\end{figure}
%
%
%
\begin{figure}[t]
\centering
\hspace{4mm} \includegraphics[width=5.cm]{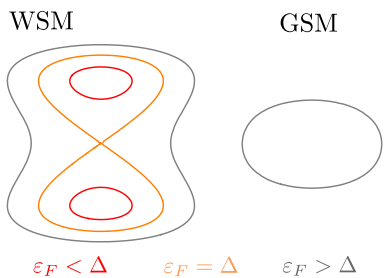}
\caption{The $k_y = 0$ cross section of the Fermi surface shown for the Weyl semimetal
and gapped semimetal case, at different Fermi energy values. For the Weyl case, the orange
curve separates the two different Fermi surface topologies. The orange line  corresponds to
the kink in the density of states.}\label{f3}
\end{figure}

%
%
\subsection{Density of states}\label{dos}
Here we calculate the density of states (DOS) for the energy dispersion from Eq.~(\ref{ham2}) for
the GSM and WSM cases.  By definition, the DOS per unit volume is,
\begin{equation}\label{g1}
  g(\e) = \frac{2}{V} \sum_{\kk} \de(\e-\e_\kk).
\end{equation}
Given the shape of the dispersions in Eq.~(\ref{ham2}), the sum is changed into an integral in a cylindrical
coordinate system, by introducing the variables $ \vo^2 = (\h v_x k_x)^2 + (\h v_y k_y)^2 $ and $z= \h c k_z$,
\begin{eqnarray}\label{g2}
 && \hspace{-10mm} g_\nu(\e) =\frac{2}{(2\pi)^3} \frac{1}{\hbar^3}  \frac{1}{v_xv_yc}\int \hspace{-0mm}\vo \, d\vo  \int_0^{2\pi} \hspace{-2mm} d\varphi \times \nonumber \\
 && \hspace{10mm} \int \hspace{-0mm}  dz \, \delta \left(\ee - \sqrt{\vo^2 + ( z^2 +\nu \D)^2}\right).
\end{eqnarray}
First, the delta function in Eq.~(\ref{g2}) is decomposed with respect to the $z$ variable into a sum,
\begin{equation}\label{g3}
 \de (...) = \sum_{z_0} \de(z-z_0) \left|\frac{ \sqrt{ \vo^2 + ( z_0^2 +\nu  \D)^2}} {2z_0( z_0^2 +\nu  \D)} \right|.
\end{equation}
Here, $z_0$ are the four roots of the argument within the $\delta$ function: $z_0 = \pm \sqrt{ \pm \sqrt{\ee^2 - \vo^2} -\nu \D}$.
Due to the absolute value, the outer set of $\pm$ points only brings a factor of $2$ in Eq.~(\ref{g2}). The $p=\pm 1$
under the square root is relevant for further evaluation, as it will determine the upper limit of the integration for $\vo$.
Inserting Eq.~(\ref{g3}) in Eq.~(\ref{g2}), and by noticing that $\e =  \sqrt{ \vo^2 + ( z_0^2 +\nu \D)^2}$, we have
\begin{eqnarray}\label{g4}
&& \hspace{-0mm} g_\nu(\e) =\frac{1}{2\pi^2} \frac{1}{\hbar^3} \frac{\e}{v_xv_yc} \sum_p \! \int \hspace{-1mm} \frac{\vo \, d\vo }{ \sqrt{\e^2-\vo^2} \sqrt{ p \sqrt{\e^2-\vo^2} -\nu \Delta } }.
\nonumber \\
\end{eqnarray}
The upper limit of the $\vo$ integration is determined by the condition that the expression under the square root in
Eq.~(\ref{g4}) be positive.  The first obvious constraint is $\vo < \e$, and the second depends on the sign $p$ and
on the type $\nu$.  

We solve the WSM case ($\nu = -1$) first. For $p=1$, the subroot expression is well defined if
$\vo < \e$.  For $p=-1$, we have two additional constraints. If $\e < \D$ then $0 < \vo < \e$, or else if $\e > \D$,
then $\sqrt{\e^2 -\D^2} < \vo < \e$.  The integral in Eq.~(\ref{g4}) for the WSM case with the constraints on $\vo$ can
be most simply written by introducing the variable $u = \sqrt{\e^2 -\D^2}$.  Then
\begin{eqnarray}\label{g5}
 && \hspace{-0mm} g_W(\e) = \frac{1}{2\pi^2} \frac{1}{\hbar^3} \frac{\e}{v_xv_yc} \times  \nonumber \\
 && \left( \int_0^{\e} \hspace{-1mm} \frac{ du }{ \sqrt{u+\D} } + \int_0^{\e} \hspace{-0mm} \frac{  \Theta(\D - \e) \, du  }{ \sqrt{-u+\D} }
 +\int_0^{\D} \hspace{-0mm} \frac{ \Theta(\e -\D)\,  du }{ \sqrt{-u+\D} } \right)
\nonumber \\
\end{eqnarray}
where $\Theta$ is the Heaviside step function. If we introduce an auxiliary function $\G(\e,\D)$,
\begin{equation}\label{g55}
 \G(\e,\D) =   \e \sqrt{\e - \D},
\end{equation}
and the unit $g_0$ as
\begin{equation}\label{g555}
 g_0 = \frac{1}{\pi^2 \h^3} \frac{1}{v_x v_y c},
\end{equation}
we can write the final result for DOS,
\begin{eqnarray}\label{g6}
 && \hspace{-4mm} g_W(\e) = g_0 \Big[ \G(\e,-\D)\, \Theta(\e-\D) \nonumber \\
 && \hspace{4mm} + \,\,  \big( \G(\e,-\D) + \G(-\e,-\D) \big) \, \Theta(\D -\e)  \Big].
\end{eqnarray}

The GSM case ($\nu=1$) follows similarly. Inspecting the subroot function in Eq.~(\ref{g4}), we see that
$p=-1$ makes the subroot expression negative and so we discard it. On the other hand, $p=+1$ restricts
$\vo$ to  $0< \vo < \sqrt{\e^2 -\D^2}$.  From the upper limit we conclude that $\e > \D$.  Using the same
substitution as in the WSM case, we have
\begin{eqnarray}\label{g7}
 \hspace{-0mm} g_G(\e) = \frac{1}{2\pi^2} \frac{1}{\hbar^3} \frac{\e}{v_xv_yc} \int_\D^{\e} \hspace{-0mm} \frac{ du }{ \sqrt{u+\D} },
\end{eqnarray}
which can be evaluated explicitly,
\begin{eqnarray}\label{g8}
 \hspace{-0mm} g_G(\e) = g_0 \,  \G(\e,\D)\,    \Theta(\e-\D).
\end{eqnarray}

It is interesting to notice that the low energy limit, $\e \ll \D$, of $g_W(\e)$ reduces to the 3D Dirac (3DD) case,
\begin{equation}\label{g9}
 g_W(\e \ll \D)  = g_0 \frac{\e^2}{\sqrt{\D}}.
\end{equation}
The three densities of states, Eqs.~(\ref{g6}), (\ref{g8}) and (\ref{g9}) are shown in Fig.~\ref{f2}.
The DOS in Eq.~(\ref{g9}) is twice the value of a single Dirac cone since in the low-energy Weyl picture,
there are two equal contributions of the Weyl points to the total DOS.  This can be  seen from Fig.~\ref{f3},
where the Fermi surface is shown for the WSM and GSM scenarios.  When Fermi energy is below $\D$, 
Lifshitz transition takes place and the WSM Fermi
surface contains two electron pockets which begin to merge at the Fermi energy $\e_F = \D$. This energy corresponds
to the van Hove discontinuity in the DOS, seen as a kink in Fig.~\ref{f2}.

The density of states for the zero gap case is most easily obtained by setting $\D = 0$ in
Eq.~(\ref{g8}). This gives
\begin{equation}\label{g10}
  g(\e)  =  g_0 \e^{3/2}.
\end{equation}
Notice that the above value of the DOS is the high energy $\ee \gg \D$ limit of Eqs.~(\ref{g8}) and (\ref{g6}).

%
%
\section{Zero-temperature dc quantities}\label{dc-tran}
Having evaluated the DOS, we can proceed to calculate the often used transport quantities: the total concentration
of conduction electrons $n$; the effective concentration of conducting electrons $n_{\al}$; the resistivity $\vo_{\al}$;
and the electron mobility $\mu_{\al}$. All calculations in this Section are performed for $T=0$ for both the GSM
and WSM cases.

%
%
\begin{figure*}[t]
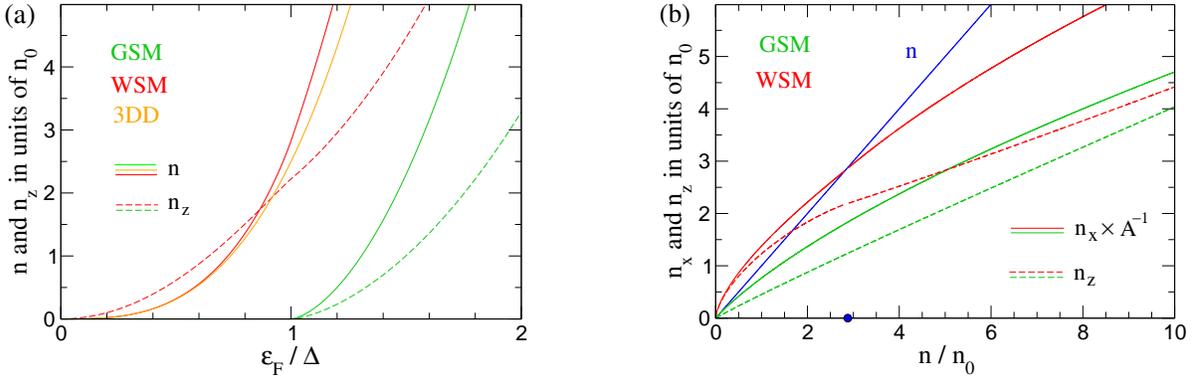

\includegraphics[width=6.99cm]{n-ovi.eps}
\hspace{1.5cm}
\includegraphics[width=7.05cm]{n-ovi-n.eps}
\caption{(a) The total electron concentrations for the gapped semimetal, Eq.~(\ref{n3}), Weyl semimetal, Eq.~(\ref{n4}),
and a low-energy limit of Weyl semimetal, which is a 3DD, Eq.~(\ref{n10}), are shown by the full lines. For the first two
cases, the effective concentration in $z$ direction is also shown by dashed lines, as given by Eqs.~(\ref{n12}) and (\ref{n13}).
All concentrations are plotted in units of $n_0$, Eq.~(\ref{n0}).
(b) The effective concentration for gapped semimetal and Weyl semimetal cases, in the $x$ and $z$ directions, $n_x$ and
$n_z$, respectively, plotted as a function of the total concentration $n$.  The $n_x$ concentrations are multiplied by
$A^{-1} \approx 0.001$ in order to fit onto the same figure. The blue dot represents the concentration $n=2\sqrt{2}n_0$
at which the states of the WSM are filled up to $\e_F = \D$, corresponding to the orange line in Fig.~\ref{f3}.}
\label{f33}
\end{figure*}
\subsection{Total electron concentration \boldmath $n$ \unboldmath}
The total carrier concentration $n$ is defined in the usual way,
\begin{equation}\label{n1}
 n = \sum_{\kk \s} f_\kk  =  \int g(\ee) f(\e,\mu)  d\ee,
\end{equation}
where the summation over bands is implicitly assumed.  At $T=0$ the Fermi-Dirac function  is $f(\e,\mu) = \Theta(\e_F -\e)$,
and it simply modifies the upper integration limit.  In integrating Eq.~(\ref{n1}) with the DOS as defined in the previous
section, we define a second auxiliary function $\N(\e,\D)$,
\begin{equation}\label{n2}
 \N(\e,\D) =   \left( 3\e + 2\D \right) \left(\e -\D \right)^{3/2}.
\end{equation}
In this way, we are able to write the total concentration of electrons in the GSM case as
\begin{equation}\label{n3}
  n_G(\e_F)  = \frac{2}{15} g_0 \N(\e_F,\D) \Theta(\e_F-\D),
\end{equation}
and similarly for the WSM case,
\begin{eqnarray}\label{n4}
&& \hspace{-6mm} n_W(\e_F)  = \frac{2}{15} g_0 \big [  \N(\e_F,-\D)\Theta(\e_F-\D) \nonumber \\
&& \hspace{-1mm} + \,  \big(  \N(\e_F,-\D) - \N(-\e_F,-\D) \big) \Theta(\D-\e_F) \big].
\end{eqnarray}
The total concentration is plotted in Fig.~\ref{f33}(a) (full lines), for the GSM and WSM cases as a function of
$\e_F/\D$, in units of $n_0$.

\subsection{Effective electron concentration \boldmath $n_\al$ \unboldmath}
The effective concentration of the conduction electrons is a direction dependent variable defined as \cite{ashcroft, gruner}
\begin{equation} \label{n5}
  n_{\al} = -\frac{1}{V}\sum_{\kk \s} m_e (v_{\al\kk})^2 (\partial f_{\kk}/ \partial \e_{\kk}).
\end{equation}
Here, $\al$ is a Cartesian component, $m_e$ is the electron bare mass and $v_{\al \kk} = (1/\h)\,\partial \e_{\kk}/ \partial k_{\al}$
is the electron group velocity.  At $T=0$,  $\partial f_{\kk} /\partial \ee_{\kk} = -\de(\e_\kk - \e_F)$, which excludes all states
except those at the Fermi level. The expression in Eq.~(\ref{n5}) forms a part of the Drude formula,
\begin{equation} \label{n6}
  \sigma_{\alpha }(\omega) = \frac{ie^2}{m_e} \frac{n_{ \al}}{ \omega  + i/\tau},
\end{equation}
where $n_{\al}$ defines the Drude spectral weight related to the plasmon frequency, which is most easily seen in a
reflectivity  measurement.  A common feature of the dispersions in Eq.~(\ref{ham2}) is the similar shape of their electron
velocity in the $\al = (x,y)$ direction, $v^\nu_{\al \kk} = \h v_\al^2 k_\al /\e^\nu_\kk$.  We have inserted this velocity
in Eq.~(\ref{n5}), so we can  evaluate $n^\nu_{x}$ for GSM and WSM case using the approach outlined in Sec.~\ref{dos}.
The result for the GSM is
\begin{equation}\label{n8}
 n^G_{x}(\e_F) = \frac{m_ev_x^2}{\D} \frac{\D}{\e_F} n_G(\e_F),
\end{equation}
and similarly for the WSM,
\begin{eqnarray}\label{n9}
 n^W_{x}(\e_F) =\frac{m_ev_x^2}{\D} \frac{\D}{\e_F} n_W(\e_F).
\end{eqnarray}
Both concentrations, Eqs.~(\ref{n8}) and (\ref{n9}), have the same high energy limit, when $\e_F \gg \D$.
For energies below $\D$, only $n^W_{x}$ remains finite,
\begin{equation}\label{n10}
 n^W_{x}(\e_F) = \frac{2}{3\pi^2\hbar^3} \frac{m_ev_x^2}{v_xv_yv_z} \e^2_F,
\end{equation}
and gives the same result as found for the 3D Dirac dispersion \cite{AshbyPRB14,Tabert2016} once we substitute
Eq.~(\ref{ham00}) in Eq.~(\ref{n9}).  The $\al =y$ case is obtained by a simple exchange  $x \to y$  in
Eqs.~(\ref{n8}) and (\ref{n9}).

A different behavior is anticipated for the $n^\nu_{z}$, primarily because of the different velocity
dependence $v^\nu_{z\kk} = 2\h c^2 k_z (\h^2c^2k_z^2 +\nu \D)/\e^\nu_\kk$.
Solving for $n_z$ calls for the definition of yet another auxiliary function,
\begin{equation}\label{n11}
 \M(\e,\D) = \left(15\e^2 +12\e \D + 8\D^2 \right) \left( \e -\D \right)^{3/2},
\end{equation}
which then yields
\begin{equation}\label{n12}
 n^G_{z}(\e_F) = \frac{4}{105}g_0\frac{m_ec^2}{\e_F}  \M(\e_F,\D) \Theta(\e_F-\D),
\end{equation}
and
\begin{eqnarray}\label{n13}
 && \hspace{-5mm} n^W_{z}(\e_F) =   \frac{4}{105}g_0\frac{m_ec^2}{\e_F} \Big[  \M(\e_F,-\D)\Theta(\e_F-\D)  \nonumber \\
 && \hspace{-3mm} + \, \big(\M(\e_F,-\D)- \M(-\e_F,-\D) \big)\Theta(\D -\e_F) \Big].
\end{eqnarray}
From Table~\ref{tabla} we see that $m_ec^2 = 1/4$, allowing us to plot both concentrations, Eqs.~(\ref{n12}) and
(\ref{n13}), in Fig.~\ref{f33}(a) in units of $n_0$ as a function of the ratio $\e_F/\D$.  What we see from
Fig.~\ref{f33}(a) is that both the total and the effective electron concentrations are very similar in shape for the
GSM and WSM at Fermi energies above  $\e_F= \D$, where the $n^\nu > n^\nu_z$. This trend is reversed for low Fermi
energies, $\e_F < \D$, where only the WSM concentrations remain finite.
In addition, the effective electron concentration for the Weyl case, Eq.~(\ref{n13}), has a weak hump at
$\e_F = \D$. This is produced by a kink in the DOS. The effective concentrations for the gapped, Eq.~(\ref{n8}),
and the Weyl case, Eq.~(\ref{n9}), in comparison to the total carrier concentrations, Eqs.~(\ref{n3}) and
(\ref{n4}), are $A = m_ev_x^2/\D = 930$ times larger if we take the values from the Table \ref{tabla}.
The parameter $A$ is used in plotting the concentrations in Fig.~\ref{f33}(b).
To that end, we introduce a unit of concentration $n_0$ for the $\D \neq 0$ cases:
\begin{equation}\label{n0}
 n_0= \frac{2}{15} g_0 |\D|^{5/2}.
\end{equation}
This unit has a value of $n_0 =  3.17 \times 10^{14} \, {\rm{cm}}^{-3}$ if the experimental values from
Table~\ref{tabla} are used.  Experimentally, it is natural to express the transport quantities as functions
of the doping or the total carrier concentration $n$.  This procedure is carried out numerically by expressing
$\e_F/\D$ as a function of the total concentration, Eq.~(\ref{n3}) for the WSM and Eq.~(\ref{n4}) for the GSM,
and then inserting this into the effective concentrations, Eqs.~(\ref{n8}) through (\ref{n13}).

Figure~\ref{f33}(b) shows the effective GSM and WSM carrier concentrations, $n^\nu_\al$, as a function of the
total carrier concentration $n^\nu$. The WSM case (red lines) is visibly different form the GSM case (green lines).
Through this difference we might obtain insight on how to distinguish the GSM case from the WSM case, at zero
temperature, based on the resistivity anisotropy. This is done in the following Section.

The zero gap case follows trivialy from Eq.~(\ref{n3}) which, after setting $\D = 0$,  gives the total concentration
\begin{equation}\label{n14}
 n(\e_F) = \frac{2}{5}g_0 \e_F^{5/2}.
\end{equation}
Setting $\D = 0$ may also be applied to all other effective concentrations.

\subsection{Mobility and resistivity}\label{otpor}

%
%
\begin{figure}[t]
\includegraphics[width=7.2cm]{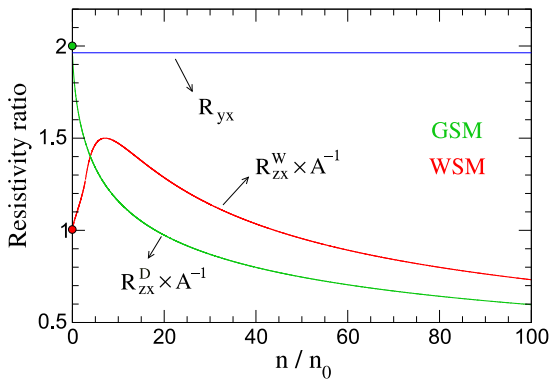}
\caption{Resistivity anisotropy defined as $R_{\alpha\beta}=\varrho_\alpha / \varrho_\beta$ and evaluated
at $T=0$ as a function of total carrier concentration for gapped semimetal and Weyl semimetal cases.
While $R_{xy}$ is constant, $R^G_{zx}$ strongly increases at low carrier concentration, and $R^W_{zx}$ has a peak.
Note that $R^G_{zx}$ and  $R^W_{zx}$ are scaled down by a factor of $1/A =1/930$.}
\label{f333}
\end{figure}

The conduction electron mobility $\mu$ is defined through the following relation \cite{mahan},
\begin{equation} \label{n14}
 \s^\nu_\al(0) = e\mu^\nu_\al n^\nu .
\end{equation}
Through comparison with Eq.~(\ref{n6}) we conclude
\begin{equation}\label{n15}
 \mu^\nu_\al = \frac{e\tau}{m_e} \frac{n^\nu_{\al}}{n^\nu }.
\end{equation}
Based on the results  for WSM and GSM cases  for in-plane effective concentrations, $\al = (x,y)$, we have,
\begin{equation}\label{n16}
 \mu^\nu_\al = \frac{e\tau}{m_e} \frac{m_ev_\al^2}{\D} \frac{\D}{\e_F}.
\end{equation}
The large ratio $A=930$ is key in the above expression, meaning that a very large intralayer carrier
mobility in ZrTe$_5$, reaching up to $0.45\times 10^{6}$ cm$^2/$(Vs), is related to a high Fermi velocity,
(Table~\ref{tabla}).

For the $z$ direction, the limit of $\e_F \gg \D$ gives the identical mobility for the WSM and GSM cases:
\begin{equation}\label{n17}
  \mu^\nu_z  = \frac{5}{14} \frac{e\tau}{m_e}.
\end{equation}
For the GSM, $\mu^G_z = 2ec^2\tau = e\tau/m^*$ when the Fermi level $\e_F$ hits just above $\Delta$.
This is a usual result for a parabolic like dispersion with an effective mass $m^*$, but interestingly
it comes with a different numerical prefactor than the high energy limit of carrier mobility [Eq.~(\ref{n17})].
The $\e_F \ll \D$ WSM case is
\begin{equation}\label{n18}
  \mu_z  =  \frac{e\tau}{m_e}  \frac{\D}{\e_F},
\end{equation}
which is equivalent to Eq.~(\ref{n16}) once we use the substitution in Eq.~(\ref{ham00}).

The direction-dependent resistivity anisotropy is best seen trough the resistivity ratio $R_{\al \beta}$, where
$(\al,\beta)  \in (x,y,z)$ defined from the Drude formula [Eq.~(\ref{n6})] for the GSM and WSM cases,
\begin{equation}\label{n19}
 R^\nu_{\al \beta} = \frac{\varrho^\nu_\al}{\varrho^\nu_\beta} =\frac{n^\nu_\beta}{n^\nu_\al}.
\end{equation}
The in-plane resistivity ratio is straightforward and equal for the GSM and WSM cases. Using
Eqs.~(\ref{n8}) and (\ref{n9}), and Table~\ref{tabla}, we get
\begin{equation}\label{n199}
 R_{y x}  = {v_x^2}/{v_y^2} = 1.96.
\end{equation}
This constant value is shown in Fig.~\ref{f333} in blue.

In contrast to Eq.~(\ref{n199}), the out-of-plane anisotropy $R_{z x}$ strongly depends on the total concentration
of electrons $n$.  This is seen in Fig.~\ref{f333} where $R_{z x}$ is plotted for the WSM and GSM cases as a function
of $n/n_0$.  The upper limit of the plot is $100\,n_0$.  For $\D = 3$~meV (Table~\ref{tabla}), this corresponds to a
Fermi energy of $\e^D_F = 13.4$~meV [Eq.~(\ref{n3})] in the GSM and $\e^W_F = 11.4$~meV [Eq.~(\ref{n4})] in the WSM case.
In the low concentration limit, $R^D_{z x}$ and $R^W_{z x}$ are visibly different.  While $R^G_{z x}$ decreases
monotonically from the maximal value of $2A$; $R^W_{z x}$ increases to a maximum located at $7.3\,n_0$, only
to start decreasing for larger doping.  This qualitatively different behavior of $R_{z x}$ as a function of $n$ is
the key to distinguish the GMS from the WSM in $dc$ transport, under the condition that  the samples can be
chemically or electrostatically doped.  Using expressions for effective carrier concentrations, Eqs.~(\ref{n6}),
(\ref{n8}) and (\ref{n9}),  gives  in the high concentration limit $n \gg n_0$,
\begin{equation}\label{n20}
 R^\nu_{zx}  \approx  \frac{14}{5} \frac{m_ev_x^2}{\D} \bigg[ \bigg(\frac{n}{3n_0}\bigg)^{2/5} + \nu \frac{2}{15}  \bigg]^{-1}  .
\end{equation}
This is shown in Fig.~\ref{f333}, where the splitting between the GSM and WSM follows from Eq.~(\ref{n20}).
In the opposite, low-energy limit when $\e_F \ll \D$, the resistivity anisotropy is only meaningful in
the WSM case where it is given by,
\begin{equation}\label{n21}
 R^W_{xz}  \approx \frac{v_z^2}{v_x^2}.
\end{equation}
The resistivity anisotropies containing $z$ and $y$ directions follow analogously.

In the zero-gap case, $R_{yx}$ is the same as Eq.~(\ref{n199}), while $R_{zx}$ is
\begin{equation}\label{n22}
 R_{zx} = \frac{14}{5}\frac{m_ev_x^2}{\e_F} \propto n^{-2/5},
\end{equation}
an exact result over the entire range of concentration $n$.  Contrary to the GSM and WSM cases, both of which
have finite values in the $n \to 0$ limit as seen in Fig.~\ref{f333}, the zero gap resistivity anisotropy,
Eq.~(\ref{n22}), diverges for small concentrations. This makes it a valuable indicator about the possible
nature of the ground state.

\section{Optical conductivity}\label{optika}
%
%
\begin{figure}[t]
\includegraphics[width=6.99cm]{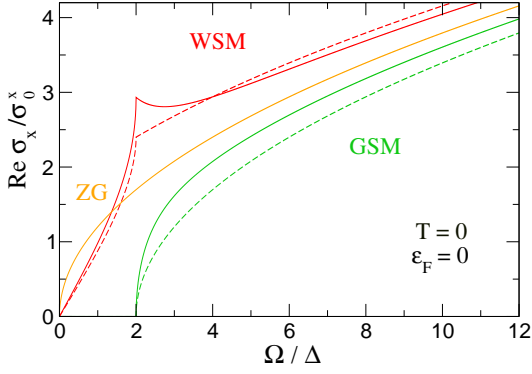}
\caption{The optical conductivity of gapped semimetal [Eq.~(\ref{op5})] and Weyl semimetal
[Eq.~(\ref{op10})] are plotted in full lines in the case of $\e_F =0$ in units of $\s_0^x$. The dashed
lines represent the optical conductivity at zero $\e_F$, but using the approximate expression Eq.~(\ref{op44})
within Eqs.~(\ref{op5}) and (\ref{op10}).  The zero gap (ZG) phase optical conductivity [Eq.~(\ref{op122})]
is shown in orange.}
\label{f4}
\end{figure}
%
%
\begin{figure*}[t]
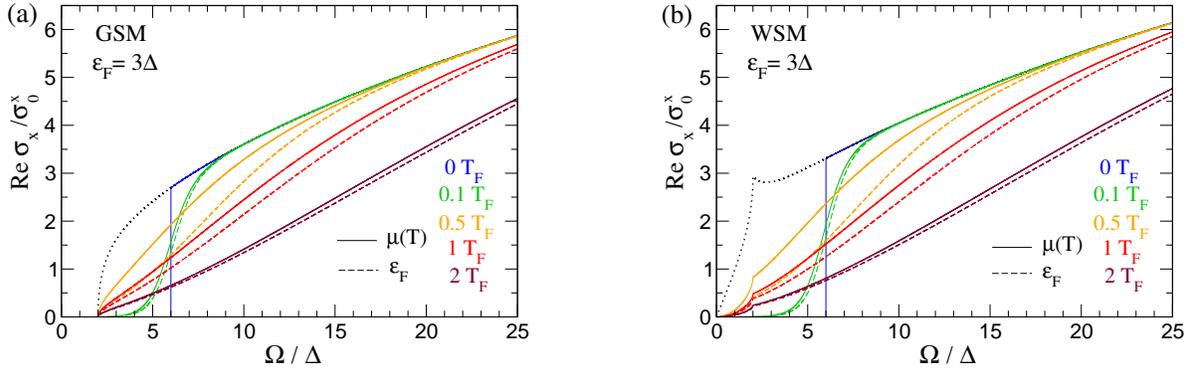

\includegraphics[width=6.99cm]{GSM-opt.eps}
\hspace{1.5cm}
\includegraphics[width=6.99cm]{WSM-opt.eps}
\caption{The $x$-component of the real part of the optical conductivity is shown for (a) the gapped semimetal
[Eq.~(\ref{op5})] is and (b) the Weyl semimetal [Eq.~(\ref{op10})]. The optical conductivity is shown at various
temperatures.  Both conductivities are given in units of $\s_0^x$, as a function of $\Omega$ which is scaled to
the gap parameter $\D$ for a particular value of Fermi energy, $\e_F = 3\D$. The full lines are optical conductivity
calculated using $\mu(T)$, while the dashed lines are calculated using constant $\mu(T) = \mu(0)=\e_F$.
The $\mathscr{F}(\Omega,T) = 1$ case is represented by dotted lines in  both panels.}
\label{f5}
\end{figure*}

When dealing with the optical response of an insulator or a semimetal, we normally use a conductivity formula
containing a phenomenological interband relaxation rate $\Gamma$.  
This interband $\Gamma$ is different from the intraband or Drude relaxation rate $1/\tau$.
In the two-band model, the interband
conductivity is \cite{Kupcic2013}
\begin{equation}\label{op01}
{\rm{Re}} \,  \s_{\al}(\Omega,T) = \frac{i\hbar}{V} \sum_{\kk \s} \frac{|J^{vc}_{\al \kk}|^2}{\ee^c_{\kk}-\ee^v_{\kk}} \frac{f^v_{\kk} - f^c_{\kk}}{\Omega-\ee^c_{\kk}-\ee^v_{\kk}+i\Gamma}+
 c \leftrightharpoons v .\\
\end{equation}
In Eq.~(\ref{op01}) we have introduced $\Omega = \hbar\om$ and the  $\al$-dependent interband current vertices
$J^{vc}_{\al \kk} $ \cite{Kupcic2016} which are calculated in Appendix \ref{apa} for the WSM and GSM cases.
Here we limit our discussion to the interband conductivity, knowing that a Drude term will always be present for a finite carrier density.

We analytically evaluate the real part of the conductivity tensor [Eq.~(\ref{op01})] in the limit $\Gamma \to 0$.
Considering only $\Omega > 0$, the above expression (\ref{op01}) becomes
\begin{equation}\label{op1}
{\rm{Re}} \,  \s_{\al}(\Omega,T) = \frac{\hbar\pi}{V} \sum_{\kk \s} |J^{vc}_{\al \kk}|^2 \frac{f^v_{\kk} - f^c_{\kk}}{\ee^c_{\kk}-\ee^v_{\kk}}\delta(\Omega - \ee^c_{\kk}+\ee^v_{\kk}). \\
\end{equation}
The Fermi-Dirac distributions in the above expression are simplified by taking into account the symmetry of
the bands $\ee^c_{\kk} = -\ee^v_{\kk}$ and the fact that the expression Eq.~(\ref{op1}) is finite only for
$\Omega = \ee^c_{\kk} -\ee^v_{\kk} $.  We can then write the distribution function as
\begin{equation}\label{op2}
\hspace{-0mm}\mathscr{F}(\Omega,T ) = f^v_{\kk} - f^c_{\kk} =  \frac{{\rm{sinh}}(\beta \Omega/2)}{{\rm{cosh}}(\beta \mu) + {\rm{cosh}}(\beta \Omega/2)}.
\end{equation}
In the $T = 0$ case, the above expression simplifies to $\mathscr{F}(\Omega,0) = \Theta(\Omega -2\ee_F)$, which describes
the suppression of the interband transitions due to the Pauli blocking.

Calculation of ${\rm{Re}} \, \s_{x}(\Omega,T)$ follows analogously to the procedure outlined in previous sections.
First we insert the interband current vertex, Eq.~(\ref{a88}), into Eq.~(\ref{op1}). The new variables are
$2\h v_x k_x = x$, $2 \h v_y k_y = y$ and $\sqrt{2} \h c k_z = z$. After the transformation into the cylindrical system,
we have
\begin{eqnarray}\label{op3}
\hspace{-20mm}&&{\rm{Re}} \, \s^\nu_{x}(\Omega,T) = \frac{1}{16\sqrt{2}\pi} \frac{e^2}{\h^2} \frac{v_x^2}{v_xv_yc} \frac{\mathscr{F}(\Omega,T)}{\Omega} \int \varrho d\vo  \int dz  \nonumber \\
 && \delta \big(\Omega - \sqrt{\vo^2 + ( z^2 +\nu 2\D )^2}\big)  \bigg(  1+ \frac{(z^2+\nu2\D )^2}{\Omega^2} \bigg).
\end{eqnarray}
The solution to Eq.~(\ref{op3}) will be facilitated by introducing yet another  auxiliary function,
\begin{equation}\label{op4}
  \DD(\Omega,\D)= \sqrt{\Omega - 2\D} \left( 1+ \frac{3\Omega^2 + 8\Delta \Omega + 32 \Delta^2}{15 \Omega^2}  \right).
\end{equation}
Here  we mention briefly some of the properties of $\DD(\Omega,\D)$.  For $\Omega$ just above  $2\Delta$, Eq.~(\ref{op4})
reduces to  $\mathscr{D}(\Omega,\D) \approx 2 \sqrt{\Omega - 2\D}$. In the opposite limit ($\Omega \gg \Delta$),
we have $\mathscr{D}(\Omega,\D) \approx (6/5)\sqrt{\Omega}$.  In all cases of interest, function $\mathscr{D}(\Omega,\D)$
can be well enough approximated by
\begin{equation}\label{op44}
 \mathscr{D}(\Omega,\D) \approx (6/5)\sqrt{\Omega - 2\D} .
\end{equation}
To simplify our optical expressions, we define the units of conductivity, $\s_0^\al$. They depend on the component
$\al \in (x,y,z)$,
\begin{equation}\label{op444}
 \s_0^\al = \frac{e^2}{8\sqrt{2}\pi \h^2} \frac{1}{v_x v_y c} (v^2_x \delta_{\al,x} + v^2_y \delta_{\al,y} + 2c^2 \delta_{\al,z} ).
\end{equation}

In continuation, we determine the optical conductivities separately for the GSM and WSM cases.

%
%
\subsection{Optical conductivity for gapped semimetal case}
The real part of the $x$-component  of the interband conductivity is given for the GSM ($\nu = +1$) case by
\begin{equation}\label{op5}
 {\rm{Re}} \, \s^G_{x}(\Omega,T) = \s_0^x \mathscr{F}(\Omega,T) \DD(\Omega,\D).
\end{equation}
Figure~\ref{f4} shows the optical conductivity determined from Eq.~(\ref{op5}) for the intrinsic case where
$\e_F=0$, in other words $\mathscr{F}(\Omega,T)=1$.  If an approximate expression shown in Eq.~(\ref{op44})
is used in Eq.~(\ref{op5}), it leads to a simplified version of the interband conductivity,
\begin{equation}\label{op6}
  {\rm{Re}} \, \s^G_{x}(\Omega,T = 0) \approx \s_0^x \frac{6}{5} \sqrt{\Omega - 2\D} \,\, \Theta(\Omega-2\e_F).
\end{equation}
This approximate result is also shown in Fig.~\ref{f4} with a dashed line, and it is rather close to the exact
expression in Eq.~(\ref{op5}).

Figure~\ref{f5} shows the real part of the optical conductivity determined from Eq.~(\ref{op5}) for various temperatures
given in units of Fermi temperature, $k_BT_F = \e_F$.  We consider two cases. In the first case, we neglect the temperature
variation of the electron chemical potential by fixing $\e_F = \mu(T=0)$.  In the second case, we include the temperature
dependence of the chemical potential $\mu(T)$, and we calculate $\mu(T)$ self-consistently from the relation Eq.~(\ref{n1})
inserted into Eq.~(\ref{op2}).
The difference between using $\mu(T)$ or $\e_F$ diminishes at $T \ll T_F$, and at  high temperatures, $T > T_F$. The reason
for this is that at low temperatures $\mu \approx \e_F$, and at high temperatures the Fermi-Dirac distribution is smeared
beyond the temperature dependence of $\mu(T)$.  Interestingly, in the intermediate temperature range where $T \sim T_F$,
the optical conductivity develops a linear-like energy dependence.  This quasi-linear optical response of ${\rm{Re}} \, \s^G_{x}(\Omega,T > T_F)$  shown in Fig.~\ref{f5} can easily be mistaken for a sign of a 3D Dirac-like band structure.

The derivation of ${\rm{Re}} \, \s^G_{y}(\Omega)$ is essentially the same, the only difference arising from the current
vertex which changes the ratio of the electronic velocities. The resulting real part of the optical conductivity is,
\begin{equation}\label{op7}
 {\rm{Re}} \, \s^G_{y}(\Omega,T) = \s_0^y\mathscr{D}(\Omega,\D)\mathscr{F}(\Omega,T).
\end{equation}
The differences are, just like in the $dc$ transport, addressed in Sec.~\ref{dc-tran}, in the $z$ direction.
This is a result of a different current vertex $J^{vc}_{z\kk}$ [Eq.~(\ref{a888})].
Introducing the fifth and final auxiliary function $\Z(\Omega,\D)$,
\begin{equation}\label{op8}
\mathscr{Z}(\Omega) =  \frac{8}{105} \frac{1}{\Omega^2}\sqrt{\Omega-2\Delta} \,(\Omega-2\D)^2(5\Omega + 2\D),  \\
\end{equation}
we can write the $z$ component of the optical conductivity in a more compact way,
\begin{equation}\label{op9}
 {\rm{Re}} \, \s^G_{z}(\Omega,T) = \s_0^z \mathscr{F}(\Omega,T)  \mathscr{Z}(\Omega,\D).
\end{equation}
Energy properties of Eq.~(\ref{op9}) are determined by the function $\Z(\Omega,\D)$, whose limit
$\Z(\Omega \gg 2\D,\D) \sim \Omega^{3/2}$ determines the high energy $z$ components of the real part of the conductivity
\begin{equation}\label{op99}
  {\rm{Re}} \, \s^G_{z}(\Omega,T) \propto \Omega^{3/2}.
\end{equation}
This function is plotted in Fig.~\ref{f6} and it is visibly different from the $xy$ plane conductivity [Fig.~\ref{f5}(a)],
which behaves as $\propto \Omega^{1/2}$. However, unfortunately the $z$-axis optical conductivity is experimentally much
less accessible.

\subsection{Optical conductivity for Weyl semimetal case}
Similar analysis applies to the WSM case. Once again using the shorthand introduced in Eq.~(\ref{op4}), the real component
of the optical conductivity along $x$ axis is,
\begin{eqnarray}\label{op10}
 &&\hspace{-6mm} {\rm{Re}}  \, \s^W_{x}(\Omega,T) =  \s_0^x \mathscr{F}(\Omega,T) \Big[  \mathscr{D}(\Omega,-\D)\Theta(\Omega-2\D)  \nonumber \\
 && \hspace{2mm} + \, \big(\mathscr{D}(\Omega,-\D)- \mathscr{D}(-\Omega,-\D) \big)\Theta(2\D -\Omega) \Big].
\end{eqnarray}
The basic features of this function are displayed in Fig.~\ref{f4}, where Eq.~(\ref{op10}) is plotted for the case that
$\mathscr{F}(\Omega,T) = 1$ and taking the full expression Eq.~(\ref{op4}), shown in full line, versus the approximation
Eq.~(\ref{op44}), shown in a dashed line.
The linearity of ${\rm{Re}}  \, \s^W_{x}(\Omega,T)$ is clearly seen for $\Omega < 2\D$. By expanding  (\ref{op10}) for
small energies $\Omega$, we have indeed,
\begin{equation} \label{op11}
  {\rm{Re}}  \,\s^W_{x}(\Omega \ll 2\D,T) \approx \frac{e^2}{6\pi \h^2} \frac{v_x^2}{v_xv_yv_z}\Omega \,  \mathscr{F}(\Omega,T),
\end{equation}
in accordance with the 3D Dirac spectrum \cite{AshbyPRB14}. At the energy $\Omega = 2\D$, a direct transition between two
hyperbolic points in the energies of Eq.~(\ref{ham2}) occurs, and manifests itself as a kink in the curve, just as it did in
the DOS. In the case of $\Omega \gg 2\D$, the optical conductivity becomes
\begin{equation}\label{op111}
 {\rm{Re}} \, \s^W_{x}(\Omega,T = 0) \approx \s_0^x\frac{6}{5} \sqrt{\Omega + 2\D} \,\, \Theta(\Omega-2\e_F).
\end{equation}
Figure~\ref{f5}(b) shows the WSM optical conductivity from Eq.~(\ref{op10}) plotted for various temperatures  $T_F$.
As in the previous calculation, the case of constant $\e_F = \mu(T=0)$ and the $\mu(T)$ have been addressed.
The $\mu(T)$ was calculated self-consistently from Eq.~(\ref{n1}).
In addition to the similar temperature dependent features like in the GSM case, we see a persistent kink at $2\D$
at all temperatures. This kink comes from the merging of the Weyl cones and the related direct transitions between
van Hove points.

The $z$ component is,
\begin{eqnarray}\label{op12}
&& \hspace{-6mm} {\rm{Re}} \, \s^W_{z}(\Omega,T) =  \s_0^z \mathscr{F}(\Omega,T) \Big[  \Z(\Omega,-\D)\Theta(\Omega-2\D)  \nonumber \\
&& \hspace{2mm} + \, \big(\Z(\Omega,-\D)- \Z(-\Omega,-\D) \big)\Theta(2\D -\Omega) \Big],
\end{eqnarray}
with the ${\rm{Re}} \,  \s_{z}(\Omega,T) \propto \Omega^{3/2}$ in the high $\Omega$ limit, like for the GSM in Eq.~(\ref{op99}).
In the low-energy limit, the above relation, Eq.~(\ref{op12}), reduces to the expression in Eq.~(\ref{op11}),
with $v_x$ replaced by  $v_z$.

%
%
\subsection{Optical conductivity for zero gap case}
In the zero gap phase, the GSM and WSM expressions from the previous two Sections reduce to the same result. Since
$\DD(\Omega,\D=0)= (6/5)\sqrt{\Omega}$, for the $\al =x,y$ components of the conductivity we get
\begin{equation}\label{op122}
 {\rm{Re}} \, \s_{\al}(\Omega,T) = \s_0^\al \frac{6}{5}\sqrt{\Omega} \, \, \mathscr{F}(\Omega,T).
\end{equation}
The above conductivity is shown in Fig.~\ref{f4}.  In a similar way, the $z$ component of the real part of the optical
conductivity is obtained by setting $\D=0$ in Eq.~(\ref{op9}), which makes Eq.~(\ref{op99}) an exact result.

%
%
\subsection{Optical conductivity anisotropy}
\begin{figure}[t]
\includegraphics[width=6.99cm]{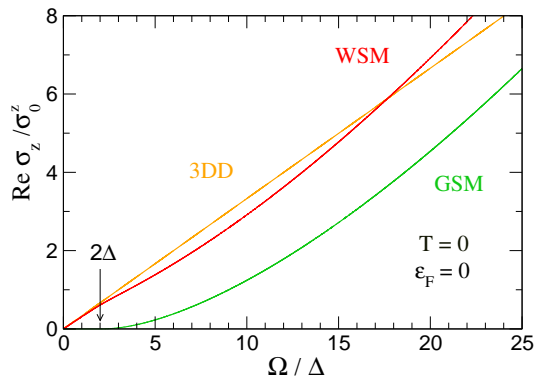}
\caption{The real part of the optical conductivity as calculated in the $z$ direction for gapped semimetal,
Eq.~(\ref{op9}), and Weyl semimetal case, Eq.~(\ref{op12}), plotted in units of $\s_0^z$. The 3D Dirac dispersion
case, Eq.~(\ref{op11}), is added for comparison. }
\label{f6}
\end{figure}

The ratio of the $x$ [Eq.~(\ref{op5})] and $y$ [Eq.~(\ref{op7})] components of the real part of the optical conductivity is
very much analogous to the analysis followed in Sec.~\ref{otpor}. For both the GSM and WSM, the optical conductivity
anisotropy is identical,
\begin{equation}\label{op13}
 \frac{{\rm{Re}} \, \s^\nu_{x}(\Omega,T)}{{\rm{Re}} \, \s^\nu_{y}(\Omega,T)} = \frac{v_x^2}{v_y^2},
\end{equation}
and given by an expression analogous to the resistivity anisotropy in Eq.~(\ref{n19}).
For the majority of anisotropic Dirac systems, the velocity ratio is $v_x/v_y \sim 1$ \cite{Morinari2009, Rusponi2010, Ryu2018},
and ZrTe$_5$ is no exception with its $v_x/v_y = 1.4$.  In some systems, this ratio was reported to be an order of magnitude
larger \cite{Park2011}.  Another equally important parameter responsible for the amplitude of the optical conductivity is the
effective mass $m^*$, hidden in $v_z$ [Eq.~(\ref{ham00})], which should be very large, $m^* \gg m_e$ for the model described
in Eq.~(\ref{ham1}) to be applicable. The effective mass $m^*$ plays a role in the following ratio which involves the $z$
component conductivity,
\begin{equation}\label{op14}
  \frac{ {\rm{Re}} \, \s^\nu_{z}(\Omega,T)}{ {\rm{Re}} \, \s^\nu_{x}(\Omega,T)} \approx  \frac{2}{3} \, \frac{\D}{m^*v_x^2} \left( \frac{\Omega}{2\D}- \nu \right).
\end{equation}
Similarly to the $dc$ case [Eq.~(\ref{n20})], because of the very large characteristic energy  $m^*v_{x}^2 \gg 1$~eV, the above ratio
is extremely small in the energy range where the model [Eq.~(\ref{ham1})] is valid.

It is worth mentioning that ${\rm{Re}} \, \s^\nu_{x,y}$ are nicely described by the approximative function, Eq.~(\ref{op44}),
compared to the exact one in Eq.~(\ref{op4}), as it can be seen from Fig.~\ref{f4}.  If we go back to the Sec.~\ref{dos}, we
may notice that ${\rm{Re}} \, \s^\nu_{x,y}$ with Eq.~(\ref{op44}) is in fact proportional to $g_\nu(\Omega,2\D)/\Omega$.
This is in accordance with the usual rule-of-thumb derivation of the optical conductivity \cite{gruner}, where the current
vertex is assumed to be a constant in Eq.~(\ref{a9}).  While this simplification  works well for the $(x,y)$ case, it utterly
fails for the $z$ direction [see Eq.~(\ref{a888})].

%
%
\subsection{Finite interband relaxation rate \boldmath $\Gamma$ \unboldmath and finite temperature effects in the GSM case}
\begin{figure}[tt]
\includegraphics[width=7.cm]{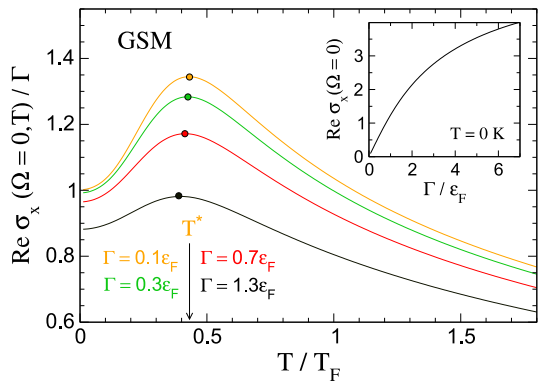}
\caption{Real part of the static interband conductivity, {${\rm{Re}} \, \s^G_{x}(\Omega=0,T)/\Gamma$, as a function of
temperature $T$, plotted for several values of interband relaxation $\Gamma$. The maximum located at $T^*$ (colored circles)
slowly shifts to lower values with increasing $\Gamma$. The inset of the picture shows ${\rm{Re}} \, \s^G_x(\Omega=0)$ as a
function of $\Gamma/\e_F$ at $T=0$. }}
\label{f6}
\end{figure}

Finite interband relaxation $\Gamma$ contribution to the  ${\rm{Re}} \, \s_{x}(\Omega,T)$  is calculated numerically from
the expression Eq.~(\ref{op1}).  Finite $\Gamma$  modifies the onset of the single particle excitation in comparison with
the analytical result in Eq.~(\ref{op3}), which then gives a nonzero value of the static interband conductivity
${\rm{Re}} \, \s_{x}(0,0)$ in the band gap region, even at zero temperatures.

The increase of the static interband conductivity can  be clearly seen in the insert of Fig.~\ref{f6}, where
${\rm{Re}} \, \s^G_{x}(0)$ is shown as a function of $\Gamma$. Deriving this functional dependence is straightforward
in the case of a 3D Dirac dispersion \cite{Tabert2016}. In the GSM case, we can find the result numerically,
\begin{equation} \label{op15}
 {\rm{Re}} \, \s_{x}(0,0) \propto \sqrt{ \Gamma} \arctan{(\sqrt{{ \Gamma}/{2\e_F}})}.
\end{equation}
The above expression shows a linear increase of the $\Omega=0$ interband conductivity for the interband damping
$\Gamma < \e_F$, and a stronger deviation for higher values of  $\Gamma$.

The temperature dependence of ${\rm{Re}} \, \s^G_{x}(\Omega=0,T)/\Gamma$ is plotted in Fig.~\ref{f6} for various
values of the interband relaxation $\Gamma$.  The strong increase of the static $T=0$ value of the conductivity is
noteworthy. This has already been addressed and is shown in the inset of Fig.~\ref{f6}.
At finite temperatures, there is a maximum located at $T^* \approx 0.4\,T_F$, which can be traced back to the
smearing of the Fermi-Dirac function  with increasing $T$. The maximum $T^*$ slowly shifts towards lower values
as we increase $\Gamma$.

This calculation is relevant in the intrinsic case, when $\e_F=0$. In the absence of a Drude component, the
interband contribution will then dominate the response.
We emphasize that the Drude component is not considered anywhere in Section~\ref{optika}, although it is present, and may be large at finite temperatures or finite carrier densities. 
%
%
\section{Conclusions}
In this article we have addressed the static and dynamic transport properties of the Weyl and gapped semimetal described
by an effective two-band model of the valence electrons.  The model implements a  linear  dispersion in the in-plane
directions and a parabolic dispersion in the out-of-plane direction, coupled to a positive band gap in the gapped case,
or a negative band gap in the Weyl case.  The transport properties in the static limit, such as the direction dependent
resistivity and mobility, are predominately influenced by large values of intralayer electron velocities. The transport
properties are similar in the Weyl and gapped cases at high values of Fermi energy. For energies lower than the band gap,
only the Weyl phase has a finite contribution, and this limit corresponds to the well-known 3D Dirac dispersion case.

In the limit of low concentrations, we show how to distinguish between Weyl phase, finite gap, or zero gap phase, using
resistivity anisotropy in the out-of-plane direction.

The interband conductivity shows a $\om^{1/2}$ dependence on photon energy in the in plane and a $\om^{3/2}$ dependence
in the out-of-plane direction for both gapped and Weyl semimetal cases.  The model  predicts that the  in-plane conductivity
anisotropy is equal to the squared Fermi velocity ratio, just like it is the case for the $dc$ transport.  The model also shows
out-of-plane conductivity anisotropy, although  proportional to $\om$, is insignificantly larger due to the comparatively large
velocity $v_x$.
The effects of a finite interband relaxation constant give a finite contribution to the interband conductivity as well as a
maximum in temperature at $T^*$, associated with the smearing of the Fermi-Dirac distribution at high temperatures and small
Fermi energies.

Finally, we showed that it is not possible to distinguish Weyl and gapped semimetal at higher temperatures and/or higher
carrier concentrations, within our effective model.  At high temperatures, both of these cases strongly resemble 3D Dirac
semimetal.
This means that the measurement of optical conductivity alone should not be used to classify the topological nature of the
ground state, if $\e_F > 2\D$ at zero temperature. Similar conclusion is valid for {\em dc} transport. If the doping is high, there is way to distinguish between Weyl and gapped semimetal.

At very low doping, $dc$ transport gives different ratios of the interlayer and intralayer resistivities for the gapped and
Weyl cases.
In the case of ZrTe$_5$, it remains an experimental challenge how to reach such low carrier concentrations.

{\em Note:} While finalizing our work, we became aware of the recent work of Wang and Li \cite{Wang2020}, whose results are in agreement with our findings for $T=0$ interband conductivity of our model.

\section{Acknowledgments}
Z.R. acknowledges helpful discussions with M.O. Goerbig.
A.~A. acknowledges funding from the  Swiss National Science Foundation through project PP00P2\_170544.
Z.~R. was funded by the Postdoctoral Fellowship of the Swiss Confederation.
This work has been supported by the ANR DIRAC3D. We acknowledge the support of LNCMI-CNRS, a member of the European Magnetic Field Laboratory (EMFL).
Work at Brookhaven National Laboratory was supported by the U.~S. Department of Energy, Office of
Basic Energy Sciences, Division of Materials Sciences and Engineering under Contract No.~DE-SC0012704.

\appendix

\section{current verticies}\label{apa}

In the general form of the  $2\times 2$ Hamiltonian
\begin{equation}\label{a1}
 H=\begin{pmatrix}
 b_{\kk} & a_{\kk} \\
 a^*_{\kk} & d_{\kk}
\end{pmatrix}
\end{equation}
the interband $L \neq\underline{L}$ current vertices can be shown to be \cite{Kupcic2016},
\begin{equation} \label{a2}
  J^{L\underline{L}}_{\al\kk} = \sum_{\ell \ell'} \frac{e}{\hbar} \frac{\partial H^{\ell \ell'}_{\kk}}{\partial k_{\alpha}}U_{\kk}(\ell,L)U^*_{\kk}(\ell',\underline{L}),
\end{equation}
where $U_{\kk}(\ell,L)$ are the elements of unitary matrix defined as $\mb{U}\hat{H}\mb{U}^{-1} = \mb{E}$
\begin{equation} \label{a3}
U_{\kk}(\ell,L) =  \begin{pmatrix}
  e^{i\varphi_{\kk}}\cos ({\vartheta_{\kk}}/{2}) & e^{i\varphi_{\kk}}\sin ({\vartheta_{\kk}}/{2})
  \vspace{2mm} \\
  - \sin ({\vartheta_{\kk}}/{2})  &  \cos ({\vartheta_{\kk}}/{2})
 \end{pmatrix},
\end{equation}
with the  definitions,
\begin{equation} \label{a4}
  a_{\kk}  =|a_{\kk}|e^{i\varphi_{\kk}}, \hspace{2mm} \tan \varphi_{\kk} = \frac{{\rm Im}\, a_{\kk}}{ {\rm Re} \, a_{\kk}},\hspace{2mm} \tan \vartheta_{\kk} = \frac{2|a_{\kk}|}{b_{\kk} - d_{\kk}}.
\end{equation}
Therefore in the general case of Eq.~(\ref{a1}), Eq.~(\ref{a2}) gives
\begin{eqnarray}\label{a5}
 && \hspace{-10mm}\frac{\hbar}{e}J_{\al \kk}^{vc} = \frac{\tan \vartheta_{\kk}}{2\sqrt{1+\tan^2 \vartheta_{\kk}}}  \frac{\partial (b_{\kk} - d_{\kk})}{\partial k_{\alpha}} \nonumber \\
 && + i |a_{\kk}|\frac{\partial \varphi_{\kk}}{\partial k_{\alpha}}
 + \frac{1}{\sqrt{1+\tan^2 \vartheta_{\kk}}} \frac{\partial |a_{\kk}|}{\partial k_{\alpha}}.
\end{eqnarray}
Now we can determine the above derivations for the Hamiltonian in Eq.~(\ref{ham1}). We obtain
\begin{equation}  \label{a6}
 \frac{ \partial |a_{\kk}|  }{\partial k_{\al}} =  \hbar\frac{v_x^2k_x\delta_{\al,x} + v_y^2k_y\delta_{\al,y}}{\sqrt{(v_xk_x)^2+(v_yk_y)^2}}
 \frac{\h^2c^2 k_z^2 + \nu \Delta }{|\ee^\nu_\kk|},
\end{equation}
and
\begin{equation} \label{a7}
\frac{ \partial \varphi_{\kk}  }{\partial k_{\al}} = \frac{v_xv_y(k_x\delta_{\al,y} - k_y\delta_{\al,x})}{(v_xk_x)^2+(v_yk_y)^2},
\end{equation}
and trivially
\begin{equation} \label{a8}
 \frac{\partial (b_{\kk} - d_{\kk})}{\partial k_{\alpha}} = 4\h^2c^2k_z\delta_{\al,z}.
\end{equation}
In the specific case of $\nu$ for the $x$ component of Eq.~(\ref{a5}),
\begin{eqnarray}\label{a88}
 &&\hspace{0mm} \frac{\hbar^2}{e^2}|J_{x \kk}^{vc}|^2 = \nonumber \\
 &&\hspace{-5mm} \frac{\h^2v_x^2}{(v_xk_x)^2+(v_yk_y)^2} \left(  v_y^2k_y^2  +  v_x^2k_x^2 \frac{ \left( \h^2c^2 k_z^2 + \nu \Delta\right)^2 }{|\ee^\nu_\kk|^2}  \right)
\end{eqnarray}
and analogously for the $\al = y$ component.
The $z$ component is rather different from Eq.~(\ref{a88}) and is
\begin{eqnarray}\label{a888}
  &&\hspace{-5mm} \frac{\hbar^2}{e^2}|J_{z \kk}^{vc}|^2 =  4\h^4 c^4 k_z^2 \frac{ {(\h v_xk_x)^2+(\h v_yk_y)^2} }{|\ee^\nu_\kk|^2}.
\end{eqnarray}
In the close vicinity of the $\Gamma$ point in the Brillouin zone $(k_x,k_y,k_z) \to 0$, and thus $\tan \vartheta_{\kk} \to 0/\D =0$. Then, inserting Eq.~(\ref{a6}) in Eq.~(\ref{a5}) for $\al = x,y$  we have
\begin{equation}\label{a9}
  |J^{vc}_{\al \kk}|^2 \approx e^2v_\al^2,
\end{equation}
while the $z$ component stays the same as Eq.~(\ref{a888}). Expanding Eqs.~(\ref{a88}) and (\ref{a888}) around Weyl points,
we again end with
\begin{equation}\label{a10}
 |J^{vc}_{\al \kk}|^2 = e^2v_\al^2,
\end{equation}
where now $\al = (x,y,z)$ with $v^2_z = 4\D c^2 $.

%
%
\bibliography{ZrTe5}

\end{document}